\documentclass[11pt,a4paper]{article} 
\pdfoutput=1
\usepackage[utf8x]{inputenc}     
\usepackage{jheppub}
\usepackage{enumerate}
\usepackage{epsfig} 
\usepackage{float} 
\usepackage[caption = false]{subfig}
\usepackage{wasysym} 
\usepackage{mathrsfs} 
\usepackage{amsfonts} 
\usepackage{amsbsy} 
\usepackage{amscd} 
\usepackage{pstricks} 
\usepackage{multirow} 
\usepackage{tikz}
\usepackage{color}
\usepackage{slashed}
\usepackage{array}
\usepackage{tablefootnote}
\usepackage{amsmath}

\usetikzlibrary{arrows,positioning,shapes.geometric} 
\usepackage[compat=1.1.0]{tikz-feynman}          
\usepackage[font=small,labelfont=bf]{caption}
\usepackage{slashed}
\usepackage{multirow}
\usepackage{tabularx}
\usepackage{soul}  
\usepackage{listings} 
\usepackage{color}

\usepackage{amsthm}

\title{
Interpretable deep learning models for the inference and classification of LHC data 
}

\author[a,b]{Vishal~S.~Ngairangbam,}
\author[b]{and Michael~Spannowsky} 
\affiliation[a]{Theoretical Physics Division, Physical Research Laboratory,\\ Shree Pannalal Patel Marg, Ahmedabad - 380009, Gujarat, India}
\affiliation[b]{Institute for Particle Physics Phenomenology, Department of Physics,\\ Durham University, Durham DH1 3LE, United Kingdom}
\emailAdd{vishal.s.ngairangbam@durham.ac.uk}
\emailAdd{michael.spannowsky@durham.ac.uk}

\abstract{ 
The Shower Deconstruction methodology is pivotal in distinguishing signal and background jets, leveraging the detailed information from perturbative parton showers. Rooted in the Neyman-Pearson lemma, this method is theoretically designed to differentiate between signal and background processes optimally in high-energy physics experiments. A key challenge, however, arises from the combinatorial growth associated with increasing jet constituents, which hampers its computational feasibility. We address this by demonstrating that the likelihood derived from comparing the most probable signal and background shower histories is equally effective for discrimination as the conventional approach of summing over all potential histories in top quark versus Quantum Chromodynamics (QCD) scenarios. We propose a novel approach by conceptualising the identification of the most probable shower history as a Markov Decision Process (MDP). Utilising a sophisticated modular point-transformer architecture, our method efficiently learns the optimal policy for this task. The developed neural agent excels in constructing the most likely shower history and demonstrates robust generalisation capabilities on unencountered test data. Remarkably, our approach mitigates the complexity inherent in the inference process, achieving a linear scaling relationship with the number of jet constituents. This offers a computationally viable and theoretically sound method for signal-background differentiation, paving the way for more effective data analysis in particle physics.
	
}
\preprint{IPPP/23/81}

\allowdisplaybreaks

\begin{document}
\maketitle
\flushbottom
\section{Introduction}
Unlike any other discipline, particle physics is blessed with a tremendous amount of highly complex data that can be theoretically understood and modelled with high accuracy. Future runs at the Large Hadron Collider (LHC) will collect even larger amounts of data, allowing for an unprecedented setup for developing novel data-analysis methods.  In recent years, machine learning, particularly deep learning frameworks applied to raw data, has been the most popular approach for efficient data processing. Notably, there are already various deep learning architectures adopted in different stages of the analysis pipeline of the ATLAS~\cite{ATL-SOFT-PUB-2018-001,ATL-SOFT-PUB-2018-002,ATLAS:2020jip,Aad:2861008,ATL-PHYS-PUB-2023-011,ATL-PHYS-PUB-2022-040,app13053282,app13010086,ATLAS:2023zcb} and CMS~\cite{Bols:2020bkb,CMS:2020poo,Pata:2750781,Cagnotta:2022hbi,Lorusso:2022xhq,May:2022lhr} collaborations. However, using deep learning techniques for classification tasks, i.e. discrimination of signal and background events, has been primarily restricted to phenomenological studies~\cite{Carrazza:2019efs,Blance:2019ibf,Anisha:2023xmh,Atkinson:2021jnj,Konar:2022bgc,Chiang:2022lsn,Lv:2022pme,Badea:2022dzb,Alhazmi:2022qbf,GomezAmbrosio:2022mpm,Freitas:2022cno}, with the community still resolving the more nuanced aspects of these powerful, yet at times black-box techniques, e.g. uncertainty estimation~\cite{louppe2016pivot,Barnard:2016qma,Bollweg:2019skg,Nachman:2019dol,Golutvin:2023fle}, reliability~\cite{DeCastro:2018psv,Araz:2021wqm,Ghosh:2021roe,Feichtinger:2021uff,Bright-Thonney:2023gdl,Metodiev:2023izu} and interpretability considerations~\cite{Chang:2017kvc,Faucett:2020vbu,Romero:2021qlf,Das:2022cjl,Komiske:2018cqr,Konar:2021zdg,Atkinson:2022uzb,Athanasakos:2023fhq,pmlr-v119-bogatskiy20a,Gong:2022lye,Bogatskiy:2022czk,hao2023lorentz}. 

The interpretation of LHC data rests on how accurately it is reproduced by simulated pseudo-data. The pseudo-data results from a theoretical calculation that factorises into the calculation of the so-called hard process, the parton shower evolution and eventually hadronisation,\footnote{For a detailed review, see for example \cite{Buckley:2011ms,Hoche:2014rga}.} i.e. the reorganisation of colour-charged partons into colour-neutral hadrons. To formalise this interpretation process, matrix element methods have been devised, either utilising calculations of the hard matrix element interfaced with transfer functions \cite{Kondo:1988yd,D0:2004rvt,Artoisenet:2010cn,Andersen:2012kn,Bury:2020ewi,Maitre:2021uaa,Dersy:2022bym,Butter:2022vkj,Heimel:2023mvw}, or even calculating the full combined hard matrix element and parton shower process from first principles. The latter method is called Shower Deconstruction~\cite{Soper:2011cr,Soper:2012pb,FerreiradeLima:2016gcz} or Event Deconstruction~\cite{Soper:2014rya,Prestel:2019neg} and constructs all possible shower histories that connect the hard process with the measured final state objects. It calculates the histories' weights and sums them up to calculate the likelihood ratio between competing hypotheses, e.g. signal process versus background process. The Neyman-Pearson Lemma proves formally that such a likelihood ratio is an ideal classifier. However, while Matrix Element methods have the advantage over machine learning techniques that do not require training, applying them to classify measured data can be computationally costly and time-consuming as the entire likelihood calculation must be redone for each measured event. The number of shower histories scales approximately factorially with the number of reconstructed jets.

To address this computation bottleneck while maintaining the high classification prowess of full-information matrix element methods, we cast the problem as a game of constructing the best shower history--ShowerMDP. It is a Markov Decision Process whose optimal policy constructs the most likely shower history corresponding to a particular process for a measured set of reconstructed objects. To showcase the method, we apply it to the discrimination of signal top-quark fat jets versus QCD-induced fat jets, noting, however, that it can be straightforwardly extended to include the full event profile, as in~\cite{Soper:2014rya,Prestel:2019neg}. Furthermore, we expand this approach by showing the potential of neural networks to learn the optimal solutions of ShowerMDP with a brute-force supervised learning approach. We call this method AlphaPS, inspired by references~\cite{Silver2016,Silver2017} on a neural agent achieving superhuman capabilities for playing the game of Go. During inference, AlphaPS scales linearly with particle multiplicity while closely matching the discrimination power of the optimal ShowerMDP policy. Thus, AlphaPS results in an exponential speedup over matrix element methods, like Shower Deconstruction, and significantly expand its applicability to high-multiplicity final states. Being based on the Lagrangian, variations of the technique could provide ways to encode task-specific physics, like data generation~\cite{Finke:2023veq,Butter:2023fov} and likelihood estimation~\cite{Rizvi:2023mws}.

The paper is organised as follows: We briefly discuss the Shower Deconstruction method and show that it maintains its prowess if one only considers the largest shower history weight  in section~\ref{sec:event_class}. In section~\ref {sec:mdp}, we give a short account of Markov Decision Processes. We formulate the problem of finding the highest weighted shower history as a Markov Decision Process in section~\ref{sec:showermdp}. The structure of the neural-assisted shower history construction and numerical results for top classification are presented in section~\ref{sec:alphaps}. Finally, we conclude in section~\ref{sec:conc}.

\section{Event classification using Shower Deconstruction}
\label{sec:event_class}

Shower Deconstruction~\cite{Soper:2011cr,Soper:2012pb,FerreiradeLima:2016gcz,Soper:2014rya,Prestel:2019neg} constructs the likelihood ratio of a measured set of particles based on the underlying QCD splittings and decay dynamics. To calculate the likelihood that a signal or a background process induced a measured final state, we construct all possible shower histories connecting the hypothesised initial state with the measured final state objects using a Feynman-diagrammatic approach. Then, we sum overall shower history weights for signal and background respectively and define the likelihood ratio $\chi_{D}(\{q^1_\mu,q^2_\mu,...q^N_\mu\})$ as
\begin{equation}
	\label{eq:chi_var} 
	\chi_{D}(\{q^1_\mu,q^2_\mu,...q^N_\mu\}) = \frac{P(\{q^1_\mu,q^2_\mu,...q^N_\mu\}|S)}{P(\{q^1_\mu,q^2_\mu,...q^N_\mu\}|B)} \quad, 
\end{equation}where $P(\{q^1_\mu,q^2_\mu,...q^N_\mu\}|S)$ and $P(\{q^1_\mu,q^2_\mu,...q^N_\mu\}|B)$  are the probabilities that the reconstructed microjets\footnote{As shower deconstruction maps measured particles to the perturbative parton evolution, the input to the algorithm should be defined in a theoretically consistent manner which is done by clustering the constituents into small radius jets and high enough transverse momentum which are referred to as microjets.} $\{q^1_\mu,q^2_\mu,...q^N_\mu\}$ originated from a signal or background processes. To get the total probability, we sum over the probabilities of all possible shower histories for a given set of particles, i.e. 
\begin{equation}
\label{eq:shower_weight} 	
\begin{split} 
	P(\{q^1_\mu,q^2_\mu,...q^N_\mu\}|S)&=\sum_{h}\; p(\{q^1_\mu,q^2_\mu,...q^N_\mu\}|h,S)\quad,\\
	P(\{q^1_\mu,q^2_\mu,...q^N_\mu\}|B)&=\sum_{h}\; p(\{q^1_\mu,p^2_\mu,...q^N_\mu\}|h,B)\quad, 
\end{split}
\end{equation}where $h$ denotes a shower history and $p(\{q^1_\mu,q^2_\mu,...q^N_\mu\}|h,H)$ denotes its associated probability for a hypothesis $H\in\{S,B\}$.

\subsection{Discrimination for top vs QCD: best history vs sum over histories}
\begin{figure}[t]
	\centering 
	\includegraphics[scale=0.29]{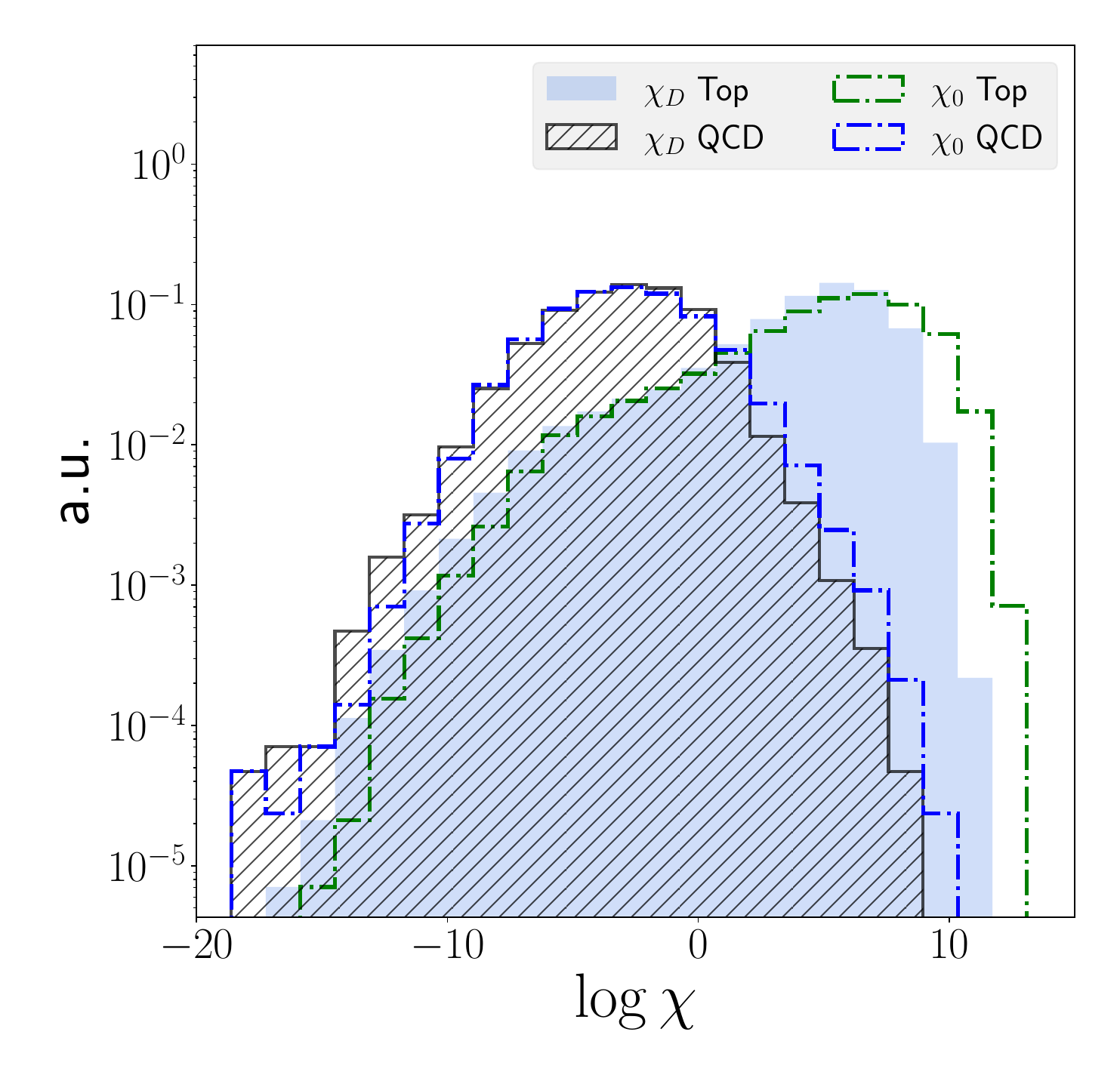}
	\includegraphics[scale=0.29]{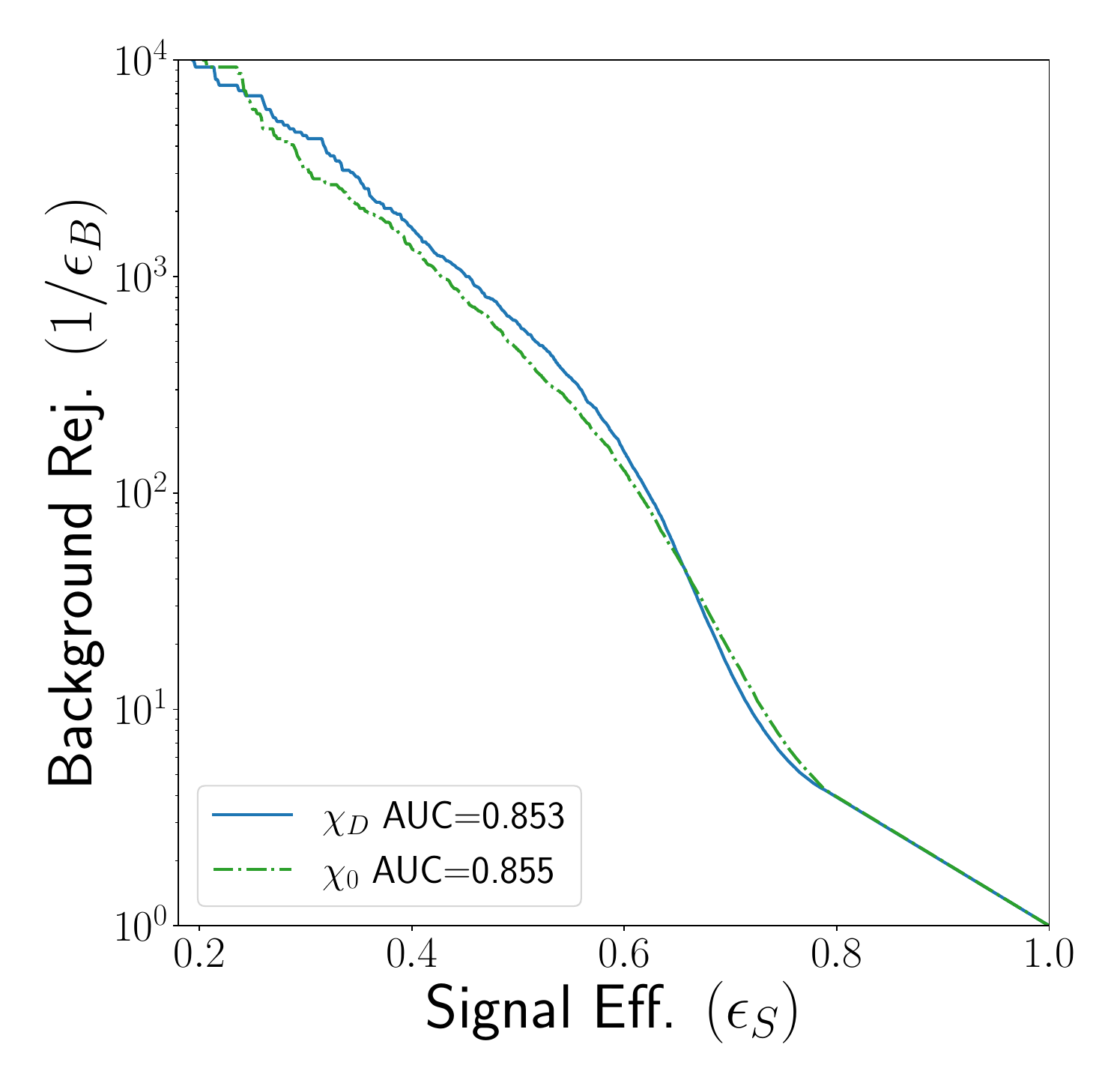}
	\caption{The figure shows normalized distributions of the log-likelihood ratio for the sum over all shower histories $\chi_D$, and the one defined with the ratio of the best shower histories $\chi_0$, for top and QCD jets. The area under the curve (AUC) for the receiver operator characteristics for top vs QCD discrimination by each variable is also displayed in the green box.}
	\label{fig:all_best_compare} 
\end{figure}

Often, we find a strong hierarchy in the shower history weights, i.e. a large number of histories are orders of magnitude smaller than a small subset of histories. Therefore, it is plausible that one can estimate $\chi_D$ to good accuracy without explicitly evaluating the entire ensemble of shower histories if the ones with large weights can be identified a priori, resulting in a gain in evaluation time. To take this argument to the extremities, we define a new likelihood ratio  
\begin{equation}
	\label{eq:chi_hist} 
	\chi_{0}(\{q^1_\mu,q^2_\mu,...q^N_\mu\}) = \frac{p(\{q^1_\mu,q^2_\mu,...q^N_\mu\}|h^{S}_0,S)}{p(\{q^1_\mu,q^2_\mu,...q^N_\mu\}|h^{B}_0,B)} \quad,
\end{equation}
where $h_0^S$ and $h_0^B$ are the histories with the highest contribution to $P(\{q^1_\mu,q^2_\mu,...q^N_\mu\}|S)$ and $P(\{q^1_\mu,q^2_\mu,...q^N_\mu\}|B)$, respectively in Eq~\ref{eq:shower_weight}. 

The left of figure~\ref{fig:all_best_compare} shows the distribution of the log-likelihood ratios\footnote{The dataset is described in Section~\ref{sec:dataset_details}. For jets with zero signal weight, we set the log-likelihood to negative one-thousand which effectively serves as negative infinity for the numerical evaluations in this work.} for top and QCD jets, showing differences visible after log scaling the $y$-axis with $\log\chi_0$ shifting consistently for both classes to the right side. The receiver-operator characteristics (ROC) curve and the area under the curve (AUC) shown on the left exhibit comparable discrimination performance. 
If an expert neural agent could construct the best shower history for a set of microjets for each process without computing all possible shower histories, there would be a massive reduction in computational time--a single shower history construction scales as $O(N)$ whereas constructing all shower histories scales naively as $O(N!)$,\footnote{Given that we are partitioning the initial set of $N$ distinct objects (identified by the four vectors) into two non-empty (except at the first division as will be seen later) and distinct subsets, the total number at each division is related to  Stirling numbers of the second kind. However, the recursive application  is more complicated as there can be repeating subtress in the full history contruction. } without any loss in the discriminatory power.

 Looking at the nature of the ROC curve, the relatively low value of the AUC (being an integrated quantity) is due to the poor background rejection at higher ranges of signal acceptance $(0.8\leq\epsilon_S\leq1.0)$. For the lower ranges of signal acceptance where a working point is usually chosen, we can see that the background rejection increases swiftly and the value of $1/\epsilon_B$ at $\epsilon_S=0.5$ is 601 (461) and at $\epsilon_S=0.3$ is 4482 (3171) for $\log\chi_D\,(\log\chi_0)$. At these working points, the algorithm's good capability is hidden by the AUC being an integrated quantity. Moreoever, as expected, $\log\chi_D$ has a better rejection than $\log\chi_0$ at these values even though it has a nominally smaller AUC. The main difference between shower deconstruction and deep learning approaches is in maximally utilising the physics which is under perturbative control in constructing an optimal observable. Being based on a matrix-element method where the perturbative part is calculated from state-of-the-art first principles, only taking into account the part of the event generation that is theoretically well understood, it shows lower accuracy than black-box neural network methods where there is no control of the information usage from different parts of the simulation pipeline. Existing studies from both ATLAS~\cite{ATLAS:2023zcb} and CMS~\cite{CMS:2020poo} show that constituent based deep learning algorithms have higher systematic uncertainties and generally scales in proportion to model complexity. Shower Deconstruction has been tried and tested at both CMS~\cite{CMS:2016tvk} and ATLAS~\cite{ATLAS:2014twa,ATLAS:2016rgb} being rather insensitive to pileup and underlying events. Consequently it has been used to set constraints on   $W'$ searches~\cite{ATLAS:2018uca}.

The rest of the paper deals with the problem of finding the highest-weighted shower history for a set of microjets and a given process. As we shall see, the natural framework for finding the best history is a Markov Decision Process (MDP); we will first outline the basic machinery of MDPs in the next section before devising one for our case.  With this formulation, neural networks can learn to construct $\chi_0$, whose value will not be explicitly dependent on the model parameters but rather through the knowledge of the learnt functions to accurately invert the parton shower dynamics, creating an interpretable framework amenable to first-principle calculations for the application of neural networks to LHC phenomenology in general. 
 
\section{A brief outline of Markov Decision Processes}
\label{sec:mdp} 

A Markov Decision Process (MDP) generalizes Markov processes\footnote{With parton shower generators based on Markov processes, it is self-evident that inverting such parton showers can be cast as a Markov Decision Process. } to situations where there is a degree of control of the environment. An agent can (at least partially) affect its evolution. They are natural frameworks for modelling real-world multiplayer games involving a discrete, possibly infinite, number of time steps. The game's rules define the environment so that an agent can learn to prioritize actions that lead to favourable outcomes in the game via interactions with the environment. Although reinforcement learning\footnote{See reference~\cite{SuttonRichardS2018Rlai} for a pedagogical introduction to reinforcement learning. } as a means to find such expert agents predate~\cite{CAMPBELL200257} the modern deep learning revolution, the availability of universal function approximators that can learn an environment's dynamics efficiently has led to breakthroughs in achieving superhuman proficiency in games as complex as Go. 

\subsection{Agent-environment interaction} 
\label{sec:agent_env} 
\begin{figure}[t]
	\centering 
	\includegraphics[scale=0.3]{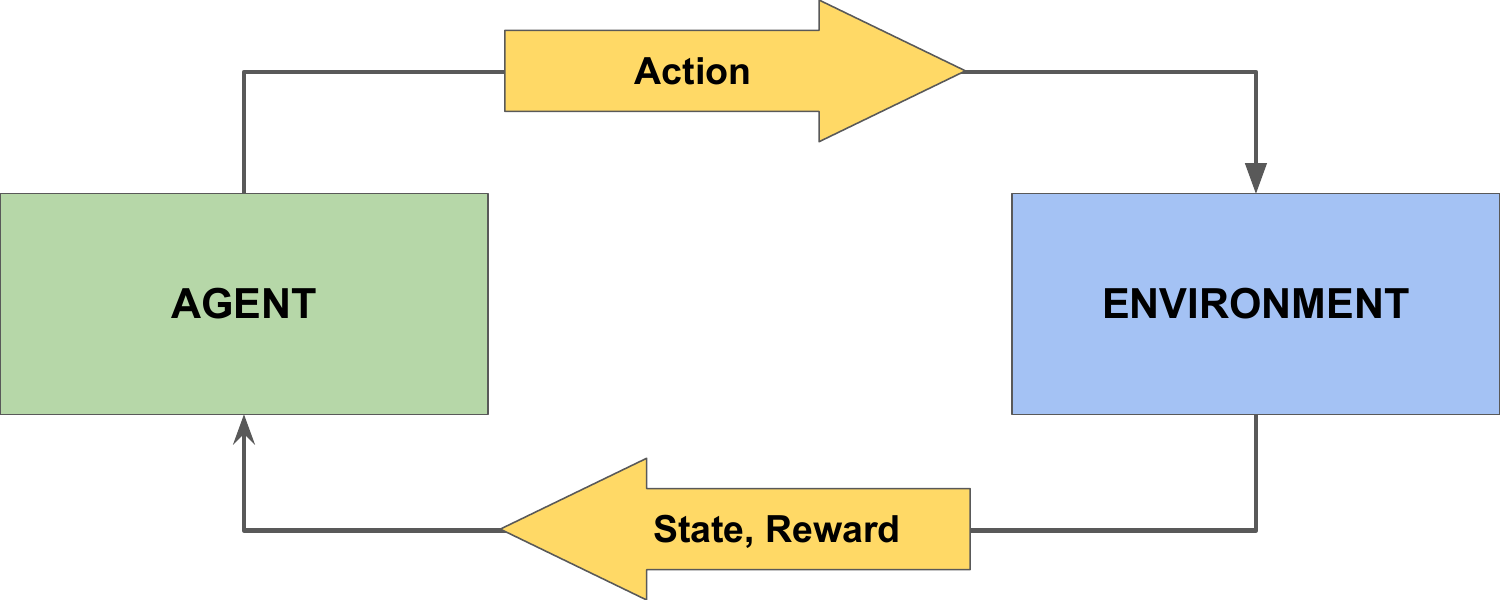}
	\caption{The figure represents the agent-environment interaction in a Markov Decision Process.}
	\label{fig:agent_env_int} 
\end{figure}

Figure~\ref{fig:agent_env_int} shows a Markov Decision Process can be encapsulated diagrammatically within an agent-environment interaction. An agent can access the environment's state with a possible set of actions. Although for many MDPs, this choice of action is not dependent on the state, it is not so for our case, and we will continue with this added complication in all our discussions. The agent acts on the environment with a particular action out of these choices, and the environment changes its state according to a predetermined transition probability. For the \emph{Markov property} to hold, this transition probability should depend only on the current state and not on previously encountered states. Based on the state transition, the environment also provides a reward to inform the agent of the suitability of the applied action. This reward is provided instantly for each step, after fixed intervals, or when the agent reaches a particular state. In this way, the agent learns to prioritize those actions that generate more rewards. The agent may have limited information about the dynamics of the environment, or the dynamics may be too complicated,  making a brute-force search for the best trajectory unfeasible. In the present case, the infeasibility of the brute-force approach with increasing microjet multiplicity motivates employing MDPs to solve shower deconstruction. 

\subsection{Policy and State Value function}  
Based on a model, an environment provides an initial state $\textbf{s}_0$ and the possible choices of actions $\mathcal{A}_0$. The agent chooses an action $\mathbf{a}_1\in\mathcal{A}_0$ according to which the environment provides a scalar reward $r_1$ and the new state $\mathbf{s}_1$. Denoting the discrete time step as $\tau$, the choices of an agent is determined by the \emph{policy} $\pi(\mathbf{s}_\tau,\mathbf{a}_\tau)$ giving the probability that an agent chooses action $\mathbf{a}_\tau$ given the state $\mathbf{s}_\tau$. The transition probability for a given state-action pair will be considered deterministic in our case and, therefore, does not warrant additional discussion. Another critical aspect of MDPs is the finiteness of the number of steps--episodic, where there is a terminal state, and non-episodic, where there is no terminal state. A terminal state $\mathbf{s}_T$ after some finite timestep $T$ is defined as one where any action produces no state transitions and zero rewards. As our design will be episodic, the following discussions will be limited to episodic MDPs.

The main goal of solving an MDP requires finding an optimal policy that maximizes the long-term reward. Suppose, a given policy $\pi$ produces the rewards $\{r_1,r_2,.....r_{T}\}$ for an episode. The total reward is simply 

\begin{equation}
	\label{eq:total_reward}
	R=\sum_{\tau=1}^{\tau=T}\; r_\tau
\end{equation}
The state value $v_\pi(\textbf{s}_\tau)$  of a state $\mathbf{s}_\tau$ is the expected total reward if one follows the policy $\pi$. The optimal policy $\pi^*$ maximizes the function $v_\pi$, denoted as $v^*$. Reinforcement learning primarily deals with finding good approximations of the optimal policy directly or indirectly by different techniques, but most are essentially based on the agent-environment interaction.

\section{ShowerMDP: the game of constructing the best shower history} 
\label{sec:showermdp} 
In the previous sections, we outlined Shower Deconstruction and Markov Decision Processes. One already notices that finding the best shower history with the maximum contribution to the total weight can be cast as a Markov Decision Process. In this section, we make this connection concrete and formulate an MDP whose optimal solution will construct the best history. This approach will be referred to as ShowerMDP in the following discussions. 
\subsection{Rules of the game: pQCD and inference at LHC} 
 
Due to the correspondence of a shower history with a Feynman diagram (although only at the soft or collinear limits or the narrow width approximation), the rules of ShowerMDP are also determined by the Standard Model Lagrangian. The Lagrangian, as we know, is the essential ingredient in Quantum Field Theory, and this approach, in principle, is not limited to the Standard Model and can incorporate BSM decays of heavy particles into coloured SM partons. 

\begin{figure}[t]
	\includegraphics[scale=0.3]{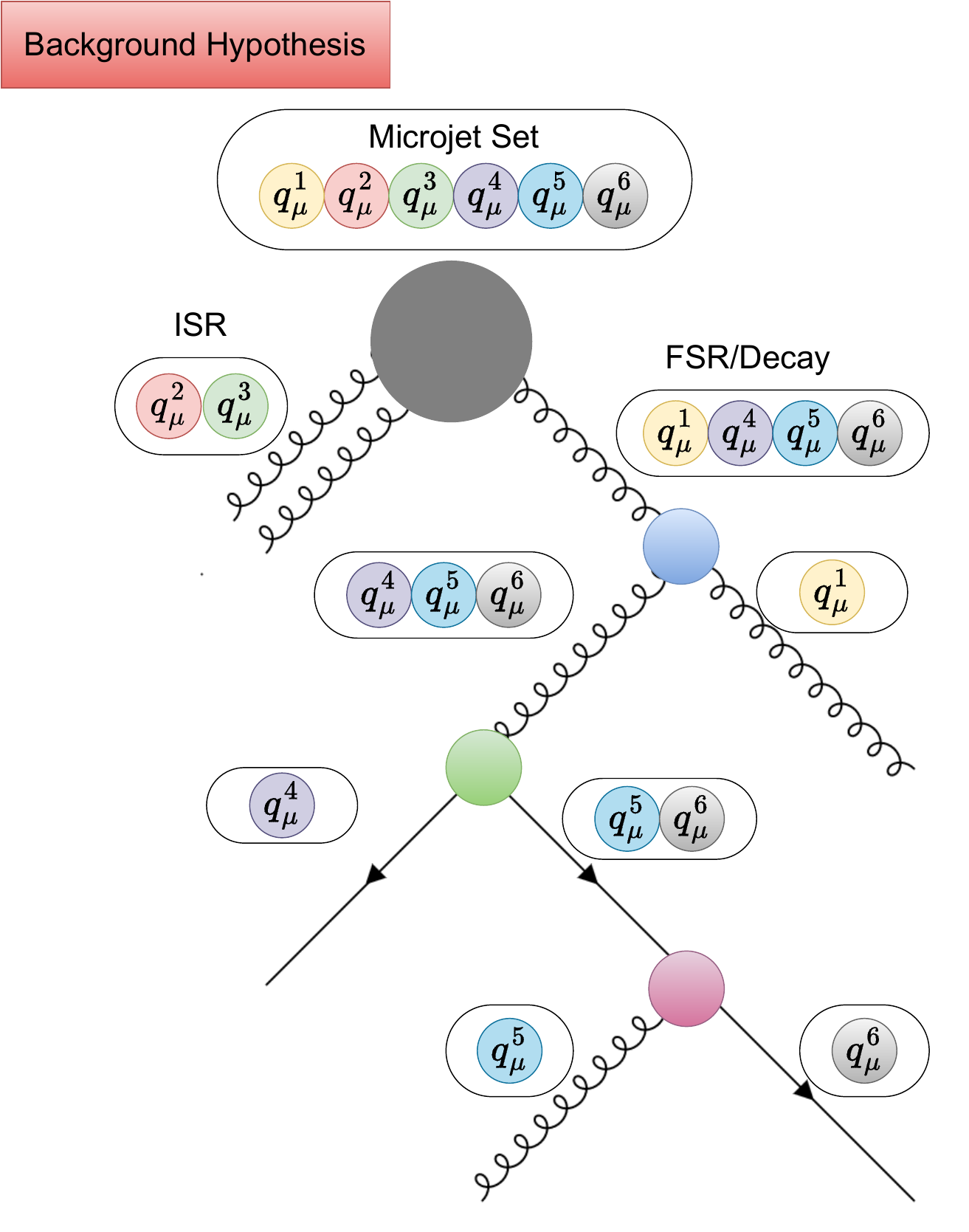}\hspace{0.1\textwidth}\includegraphics[scale=0.3]{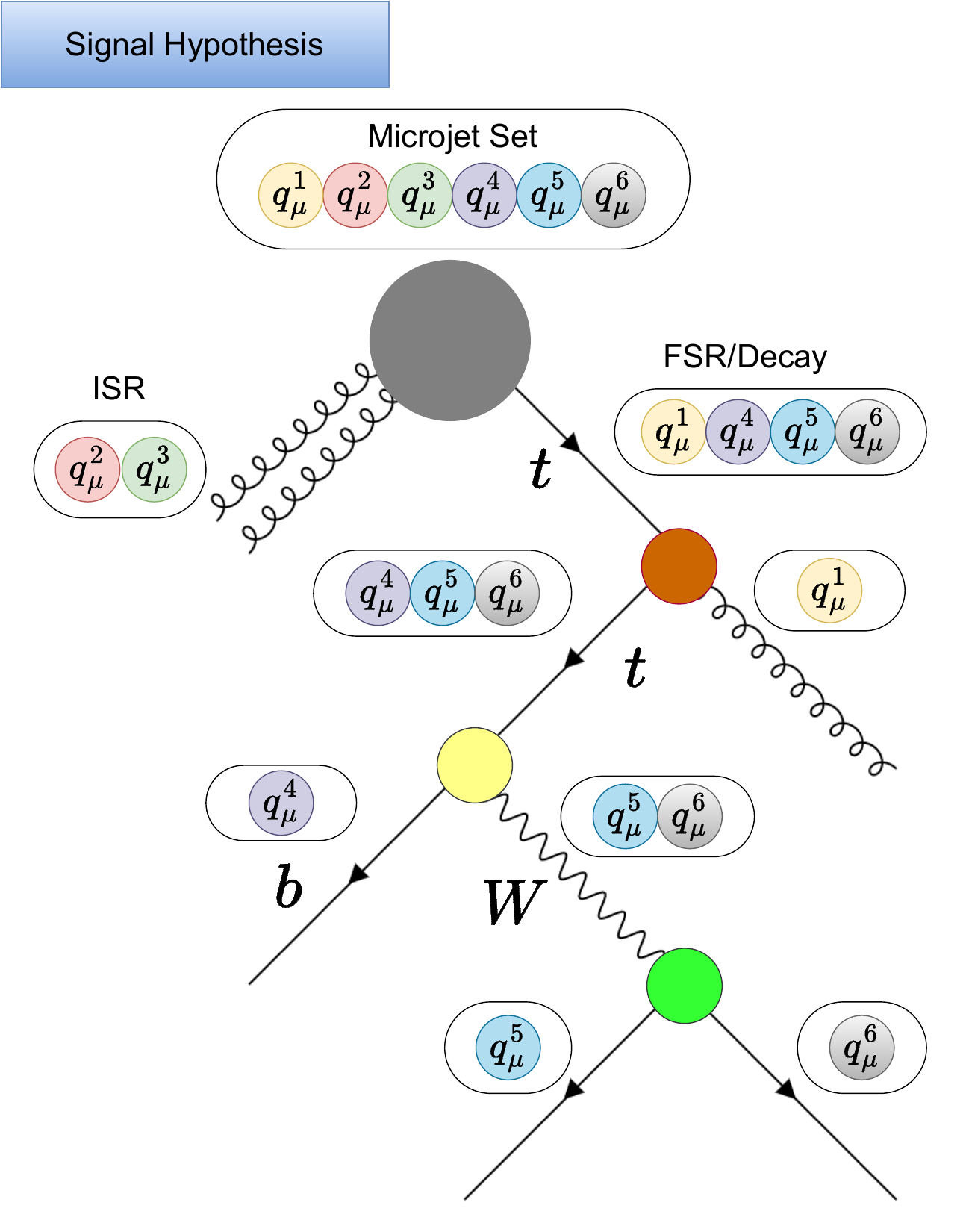}
	\caption{The figure shows a single shower history of a jet containing six microjets for the background (left) and signal (right) hypothesis.}
	\label{fig:shower_hist} 
\end{figure}

We will describe the rules by visualizing an episodic construction of the signal and background shower histories. As shown in figure~\ref{fig:shower_hist}, a shower history for a process represents a possible realization of a Feynman diagram. At the start of the game, we are provided with a set containing all the microjets of the fatjet. The agent should divide this set into two mutually exclusive subsets whose union gives the parent. The first splitting corresponds to how the microjets are divided into two mutually exclusive subsets corresponding to the initial state radiation (ISR) and underlying events (UE) contamination and those that arise from the hard parton which we call final state radiation (FSR) set. To simplify the environment, we are only concerned with the relative importance of the ISR microjets and the exact ISR weights as described in reference~\cite{Soper:2011cr} are calculated for the ISR set. 
Since the ISR contamination can be absent, we make it possible for this ISR set to be empty. Depending on the hypothesis, the flavour of the FSR set is assigned--top quark for the signal (right) and gluon (left) for the background hypothesis. 

Other than the root and the ISR edge, all other shower edges have an assigned flavour corresponding to a particle of the Standard Model. Therefore, the game's rules are the Feynman rules of the Standard Model, where the degrees of freedom depend on the splittings' energy scale. At each stage, the player splits the given microjet constituents into two mutually exclusive subsets covering the initial set. The game continues until we encounter a zero-weighted splitting or if all undivided edges contain a single microjet. The former entails a premature end of the game without reaching the goal if there are non-zero shower histories, while the latter represents a successful completion of the goal. The shower histories in figure~\ref{fig:shower_hist} illustrate completed episodes for each hypothesis using the constructed Feynman diagrams. Even though the Feynman rules are self-evident, we will describe them briefly for completeness.

Two terms in the SM Lagrangian are modelled for the top quark--the $SU(3)_C$ gauge interaction for the gluon emissions and the $SU(2)_L$ gauge interaction with the $W$ boson depicting its decay. We group the four light flavours into a single light quark category for the gluon and treat the bottom quark interaction separately. Therefore, we have three related terms in the Lagrangian at $O(\alpha_s)$--the three gluon self-interaction, $SU(3)_C$ gauge interaction of the gluon with light quark-antiquark pair, and with a pair of bottom quarks. For the top quark and the gluon, the action space is a product of the corresponding type of interaction (or flavour assignment) and the partition of the microjet set. The cardinality of the product set will represent the total number of possible action/splitting choices in the game tree. In other words, we have separate actions for each flavour assignment for the same set partition.

We will have only the multiplicity-dependent actions for the remaining flavours as we take only one representative term from the SM Lagrangian. We have the gluon emission term for the five light quarks. For the $W$ boson, we decay to a quark-antiquark pair of the same generation, grouping the two generations into a single representative term. However, there are four edge classes for the quarks besides the top-quark: quark and antiquark edges for the first two generations and bottom and antibottom edges. The multiplicity requirement of the top quark and the $W$ boson edges generate zero weights when not satisfied and is therefore not implemented as an additional rule. Consequently, we have a uniform set-partitioning action dependent only on the microjet multiplicity for all flavours. We can now give precise mathematical descriptions of the action, the state, and the reward.

\subsection{Constructing the action space} 
\subsubsection*{Set partitioning action} 
Let $S=\{p_\mu^i\; |\; i\in I\}$ be the set of microjets indexed by $I=\{1,2,....,n_e\}$ for a given shower edge of $n_e$ microjet multiplicity in the shower history. Independent of the flavour, we need to divide this mother set into two mutually exclusive subsets which cover the set. Each subset will be assigned a flavour depending on the context, which will usually differ. Even if they are the same, for instance, when a gluon splits into two gluons, the colour flow and hence the left and right colour connected partners will differ. Therefore, we must distinguish between the two subsets and designate $L$ and $R$ as the edge that inherits the mother's left- and right-color-connected partner, respectively. For the $W$ boson decay, $L$ and $R$ are defined according to which partons have a right and left color-connected partner, respectively.

Keeping the uniqueness of the two daughter edges in mind, we define the set partitioning action $\mathbf{a}_{\text{part.}}$,  for a single action edge of $n_e$ microjet constituents in the game tree as a boolean vector of $n_e$ dimensions. If a particular index of $\mathbf{a}_{\text{part.}}$ is one, then the element with that label belongs to the $L$ edge. Those with zero will then naturally belong to the $R$ edge. 

 \subsubsection*{Flavor assignment action} 
The flavour assignment action assigns the flavour of the $L$ and $R$ daughter edges. Other than the top and gluon edges, these are trivial for all other flavours. Choosing which type of splittings to follow can be cast as a classification problem where the number of classes is given by the types of splitting nodes possible for each edge. Therefore, we encode the flavour assignment for a single splitting as a one-hot encoding. For the top quark, we define 
\begin{equation} 
	\label{eq:top_flav_assign} 
	\mathbf{a}_{\text{assign.}}^{(t\;\to W\;b)}= [1,0]\quad, \text{and}\quad \mathbf{a}_{\text{assign.}}^{(t\;\to t\;g)}= [0,1] \quad.
	\end{equation} 
Similarly, for the gluon we define
 \begin{equation} 
 	\label{eq:gluon_flav_assign} 
 	\mathbf{a}_{\text{assign.}}^{(g\;\to g\;g)}= [1,0,0]\;,\;\mathbf{a}_{\text{assign.}}^{(g\;\to q\;\bar{q})}= [0,1,0]\;,\; \text{and}\quad \mathbf{a}_{\text{assign.}}^{(g\;\to b\;\bar{b})}= [0,0,1] \quad.
 \end{equation}  
 
\subsection{State definition} 
For the Markov property to hold, we need to define the state of ShowerMDP such that the choices of actions depend entirely on the current state and not on previously encountered ones. Therefore, the state must contain all the information that goes into evaluating the splitting kernel and the Sudakov factor. Here, there is a degree of dependence on the showering model. For the splitting factors, considering the color-connected partner(s) of the current edge suffices. In comparison, we need to encode the energy scale at which the edge was produced for the Sudakov factor, i.e., the mother edge's information. These features are enough if we want to generate a parton shower where the division of the edge's momentum can be continuous. However, in the present case, we are confined to discrete divisions given by the edge's microjet content. Therefore, this microjet content must also be included in the state definition. Considering a modular architecture of the neural function approximation, we call the former state the \emph{tree state} and the latter the \emph{microjet state}.   
\subsubsection*{Tree State}
The number of particles that enter the tree state definition is fixed for a given flavour. As we will be modularising the agent for each distinct flavour and hypothesis, we could, in principle, use a single vector representation containing the feature of all particles. However, we define the tree state as a point cloud for the efficient extraction and combination of information with the microjet state, which demands a point cloud representation due to the varying multiplicity.  

For any given flavour, we have at most four shower edges entering the tree state: the edge which will be split, its mother, and its colour-connected partners. Except for a gluon with a known mother in the shower history, any edge will have less than four particles in the tree state definition. Note that we do not assign the root edge as the mother of the FSR edge for both hypotheses since we are considering the ISR contamination in the fatjet and not a proper splitting in the sense of other edges in later stages of the game tree. We do not consider the tree state for the root edge since there is no information (as will be defined in Eq.~\ref{eq:tree_node_feat_edge}) in the root edge's node feature with all entries uniformly zero or one for any given microjet set.   

\begin{figure}[t]
	\centering 
	\includegraphics[scale=0.35]{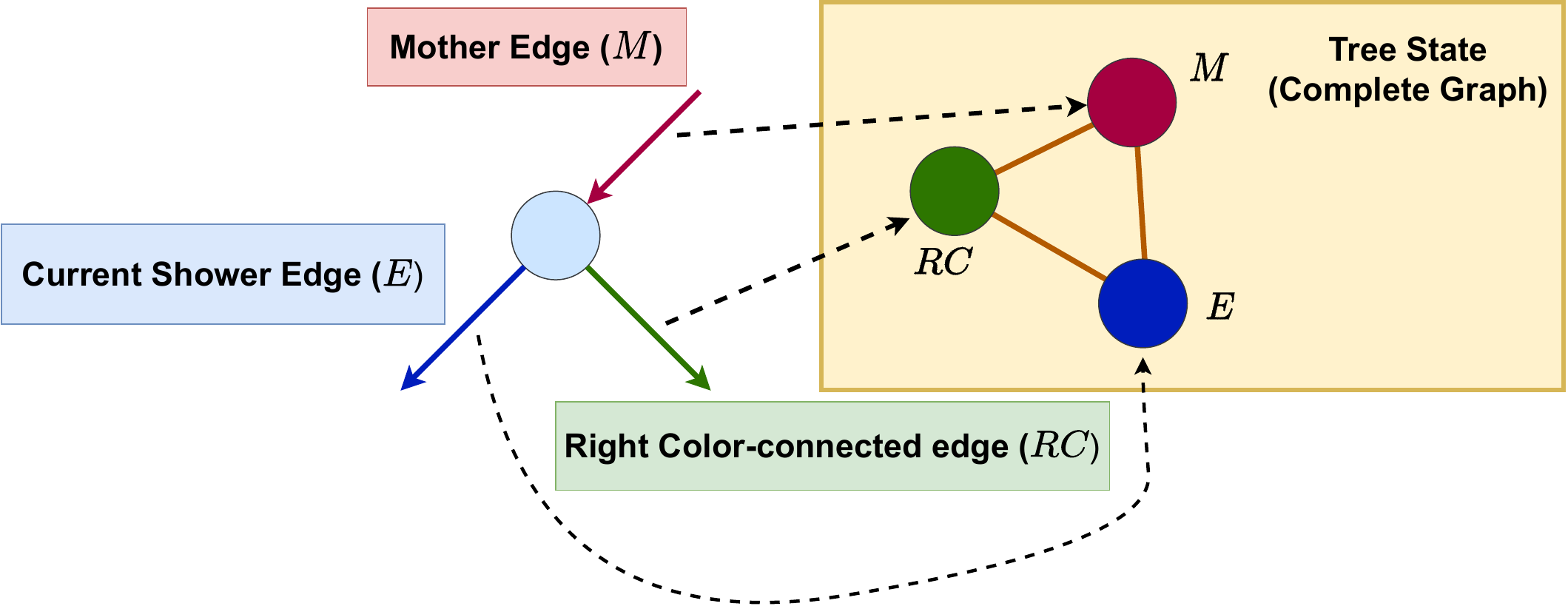}
	\caption{The figure shows a schematic representation of the mapping between the shower edges of the game  tree to the nodes of the tree state. The colors of the edges are made different to figuratively distinguish the edges in the tree state. Note that for a gluon, there will be an additional color partner in the tree graph and we will have a complete graph of four nodes in the tree state.} 
	\label{fig:tree_state} 
\end{figure}

Let us denote the shower edge the agent encounters at the particular game stage by $E$, its mother edge by $M$, the left and right color-connected edges as $LC$ and $RC$, and the jet by $J$. At this juncture, we must point out that we are dealing with a second graph when defining the tree state. Each tree state is a complete graph with a corresponding node for each of the four shower edges in the game tree. This mapping is shown in figure~\ref{fig:tree_state}.
If the flavour of the edge $E$ does not require the definition of left or right colour-connected partners, or if the mother edge $M$ is not defined at the particular stage in the shower history, they do not enter the tree state definition. For coloured partons which need colour-connected partners but are not allocated due to the incomplete structure of the shower histories, we define soft colour partners with four vectors $q_\mu^{LC}$ or $q_\mu^{RC}$ with $0.5$ GeV transverse momentum, $0.1$ GeV mass that falls outside the jet. For the left color-connected partner,  we set the pseudorapidity $\eta^{LC}=\eta_J-\frac{3}{2\sqrt{2}} R_{jet}$ , and azimuthal angle $\phi^{LC}=\phi_J-\frac{3}{2\sqrt{2}} R_{jet}$. For $q_\mu^{RC}$, we set the angles to   $\eta^{RC}=\eta_J+\frac{3}{2\sqrt{2}} R_{jet}$, and $\phi^{RC}=\phi_J+\frac{3}{2\sqrt{2}} R_{jet}$, with $\eta_J$ and $\phi_J$ denoting the pseudorapidity and the azimuthal angle of the jet.  

We define a five-dimensional node feature for a shower edge $i$ in terms of the ratios of transverse momentum $p_T^i/p_T^j$, and squared mass $m_i^2/m^2_j$; differences in rapidity $\Delta y_{i,j}$,  and azimuthal angle $\Delta \phi_{i,j}$; and an edge embedding scalar $e_i$ to distinguish between the edge's labels. 
The node feature of the shower edge $E$ is with respect to the jet
\begin{equation} 
	\label{eq:tree_node_feat_edge} 
	\mathbf{h}_E=\left[\frac{p_T^E}{p_T^{J}},\frac{m^2_E}{m^2_{J}},\Delta y_{E,J},\Delta \phi_{E,J},e_{E}\right]\quad,
\end{equation} On the other hand, the node features for $a\in \{M,LC,RC\}$ 
is defined with respect to $E$  as    
\begin{equation} 
	\label{eq:tree_node_feat_other} 
	\mathbf{h}_a=\left[\frac{p_T^a}{p_T^{E}},\frac{m^2_a}{m^2_{E}},\Delta y_{a,E},\Delta \phi_{a,E},e_{a}\right]\quad,
\end{equation}where instead of the jet momentum, we have used $p_\mu^E$. For the four shower edges, we set $e_E=1$, $e_M=-1$, $e_{LC}=-2$ and $e_{RC}=2$.

Since the top and $W$ shower edges are necessary for determining the decay probabilities, we also include each node's mass information in the tree state's node features since their mass is essential for determining the decay probabilities. The mass information for all nodes $i$ in the tree state definition is added as $m_i/100$ to keep the range of the node features uniform. 

Another information which is also relevant for the tree state is the flavour information of the edges. We supply them as the PID~\cite{Workman:2022ynf} with all four light quarks assigned the up quark's PID. Since these are arbitrarily based on convention and not on some underlying physical principle, we will utilise a neural network to learn a vector embedding of the flavour information of each node and supply the concatenated vector with $\mathbf{h}_i$ as inputs to the network processing the tree state.   

\subsubsection*{Microjet State} 
The microjet state is also defined as a point cloud, with each microjet denoted as a node. Similar to the tree state, the node feature of each microjet $i$ is defined as 
\begin{equation}
	\label{eq:micro_state}
	\mathbf{h}_i=\left[\frac{p_T^i}{p_T^{E}},\frac{m^2_i}{m^2_{E}},\Delta y_{i,E},\Delta \phi_{i,E}\right]\quad,
\end{equation}
Since we will use point-transformers for the microjet set, we will not put any additional graph construction on the microjet set. Instead, we will use the complete graph and learn the importance of the different microjets with the attention mechanism. 

\subsection{Modeling the reward}

For episodic games, the reward of each non-terminal state is determined after the end of the episode. As we will not be doing exploration-based learning,  we define the reward of each non-terminal state as the shower weight evaluated after the episode ends (either prematurely via a zero-weight splitting or after all non-ISR leaves contain exactly one microjet.)  Such a reward structure can extract the optimal shower history $p_{h_0}$ for each hypothesis.

\section{AlphaPS: solving ShowerMDP with supervised learning} 
\label{sec:alphaps} 
Having laid the groundwork, we present numerical results of the supervised training procedure used to solve the ShowerMDP, taking top-tagging as an example. We call this network \textsc{AlphaPS}, directly inspired by references~\cite{Silver2016,Silver2017} for mastering the game of Go. Our supervised approach bears similarities to \textsc{AlphaGo Fan}~\cite{Silver2016}, with the differences arising because of the variable action space, availability of an exact environment model, and the single-player nature of ShowerMDP.
The reason for reverting to a supervised training method is as follows. Unlike the application of reinforcement learning in other domains, we have a concrete benchmark to which we need to compare--the best shower history given by the brute force search, which should not be sidestepped in a proof-of-principle analysis. In other words, checking that a supervised training procedure can learn the optimal set of splittings when directly provided is necessary before we attempt a reinforcement learning procedure relying on the agent-environment interaction where the agent needs to find the best set of actions via explorations. 

Exploitation versus exploration--whether to exploit the current best actions or explore previously unexplored trajectories that may yield higher rewards, is a subject of close study in any reinforcement learning situation. Situations, where there is no exact environment model, demand a reinforcement learning procedure to search for possible optimal policies with explorations of the environment. In our case, where we have a precise ShowerMDP model, a reinforcement learning procedure is justifiable only if the exploration stage uses fewer resources than the brute force strategy. Therefore, we need to devise an efficient exploration algorithm, leading to lower computational expenditure with an efficient learning capability. This resource reduction is all the more important for extending the method to higher multiplicities.  In the present case, we discard exploitation entirely since the supervised approach can be regarded as the agent acting after a complete exploration of the environment.

\subsection{Details of the dataset}   
\label{sec:dataset_details} 

We study the ability of the neural function approximation of the optimal ShowerMDP policy for QCD (background) vs top (signal) jets at 14 TeV LHC. Parton-level dijet and top pair production processes for the background and signal were generated using \texttt{MadGraph5\_aMC@NLO (v3.5.0)}~\cite{Alwall:2014hca} in the five-flavour scheme with a minimum transverse momentum cut of 380 GeV on the leading jet and top quarks. The top quarks were decayed into the fully hadronic channel using \texttt{MadSpin}~\cite{Artoisenet:2012st}. These parton-level events were showered and hadronised with \texttt{Pythia8 (v8.306)}~\cite{Bierlich:2022pfr} along with multi-parton effects. To facilitate b-tagging within the jet, which is an important factor in the discrimination power of shower deconstruction, we keep all B-mesons stable by setting the user hook \texttt{mayDecay=off}.  All stable particles after hadronisation are clustered into jets of radius R=1.0 with the anti-$k_t$ algorithm implemented in \texttt{FastJet (v3.4.0)}~\cite{Cacciari:2005hq,Cacciari:2011ma}. We select the leading jet in an event if it is within $|\eta|<3$ and $p_T\in(400,500)$.

To keep the multiplicity under control for inputs to the brute-force extraction of shower weights, we recluster the jet constituents with the Cambridge-Aachen~\cite{Dokshitzer:1997in} algorithm and a seed radius of $R_0=0.1$, which is increased in steps of 0.01 until the clustering yields at most eight inclusive microjets with $p_T>5$ GeV. Jets with less than three microjets are not selected as they trivially yield zero signal weights under shower deconstruction. After neglecting these jets, we have 500k, 200k and 100k jets for each class in the training, validation, and test datasets, respectively. We implement bottom-tagging in these microjets by assigning a b-tag to the nearest microjet within $\Delta R<0.2$ of a B-meson.   These microjets go into the ShowerMDP. Many jets will produce zero signal weights as combinatorial realisation may not satisfy the top and $W$ mass windows. Due to the nature of the physics-transparent weight construction, these configurations are statistically more likely for QCD jets and almost negligible for top jets. We do not extract the shower history states for both hypotheses and will employ a value network to filter these jets with zero signal weights.

For jets passing the zero-weight check, we evaluate the $\log\chi_D$ and $\log\chi_0$ and extract the shower history corresponding to $\chi_0$ for both hypotheses used to train the policy network modules.

\subsection{Neural assisted construction of the best shower history}

To construct the best shower history, we learn two parametrised functions with neural networks--the value network and a composite policy network. Here, we concentrate on the essential features and expand on the architecture's detail and training methodology in appendix~\ref{app:arch}.

The value network takes the initial state $s_0$ of the root edge, which consists of the microjet state alone and is trained on all jet samples. We teach the network to predict -1 (+1) for those jets with zero (non-zero) total weight. During inference, if the network output is negative for a given jet, we set $\chi_0$ to zero, and the microjet set does not undergo any history construction. Given the complexity of the brute-force construction of all background histories and the combinatorial nature of the assigned divisions for the signal histories, which may go till the $W$ boson decay (on the third level of the tree), this step already saves a considerable computational cost, dependent, of course, on the complexity of the value network.

The policy network consists of different submodules for specific hypothesis-flavor pairs. For the background hypothesis, we combine the bottom, antibottom, quark, and antiquark flavours into a single module due to their similarity. Since the bottom quark originates from the top quark for the signal, we separate the bottom module for the signal while still using a combined module for the other three flavours.

To better handle the variable multiplicities, we use message-passing neural networks, and therefore, the network supplies a binomial probability of the microjet assignment action for each node in the microjet state. The gluon and top quark modules also have an additional softmax output for the flavour assignment actions, which has a constant dimension regardless of the microjet multiplicity. The extracted state-action pair for all non-zero histories of both classes are divided into their specific hypothesis-flavour pair, and the modules are trained independently.

We have two episodes for each hypothesis on those jets with positive predictions by the trained value network during inference. Starting from the root node, we consider the assignment of each microjet as a Bernoulli distribution given by the policy network's microjet assignment action. An action is sampled from this distribution and applied to the initial microjet set. For the top quark and gluon sets, we independently sample the flavour assignment action via the split action output of the policy. The process continues until we get a complete shower history or a zero-weighted action. The latter ends the episode prematurely. Due to the stochastic nature of the NN-assisted construction, we extract the shower weights for one thousand episodes and study the dependence of the maximum shower weights on the number of repetitions. Due to the independence of the Bernoulli probabilities, we can have instances (although relatively rare) where all of them are either very close to zero or one. A random sampling using these probabilities would not produce a correct set partition, and we set the upper limit to produce viable set partitions to one hundred trials. Suppose we cannot get a proper set partitioning action after a hundred trials. In that case, a possible action is implemented by either setting the maximum (minimum) valued probability to one (zero) when all sampled actions are zero (one).

\begin{figure}[t!]
	\centering
	\includegraphics[scale=0.29]{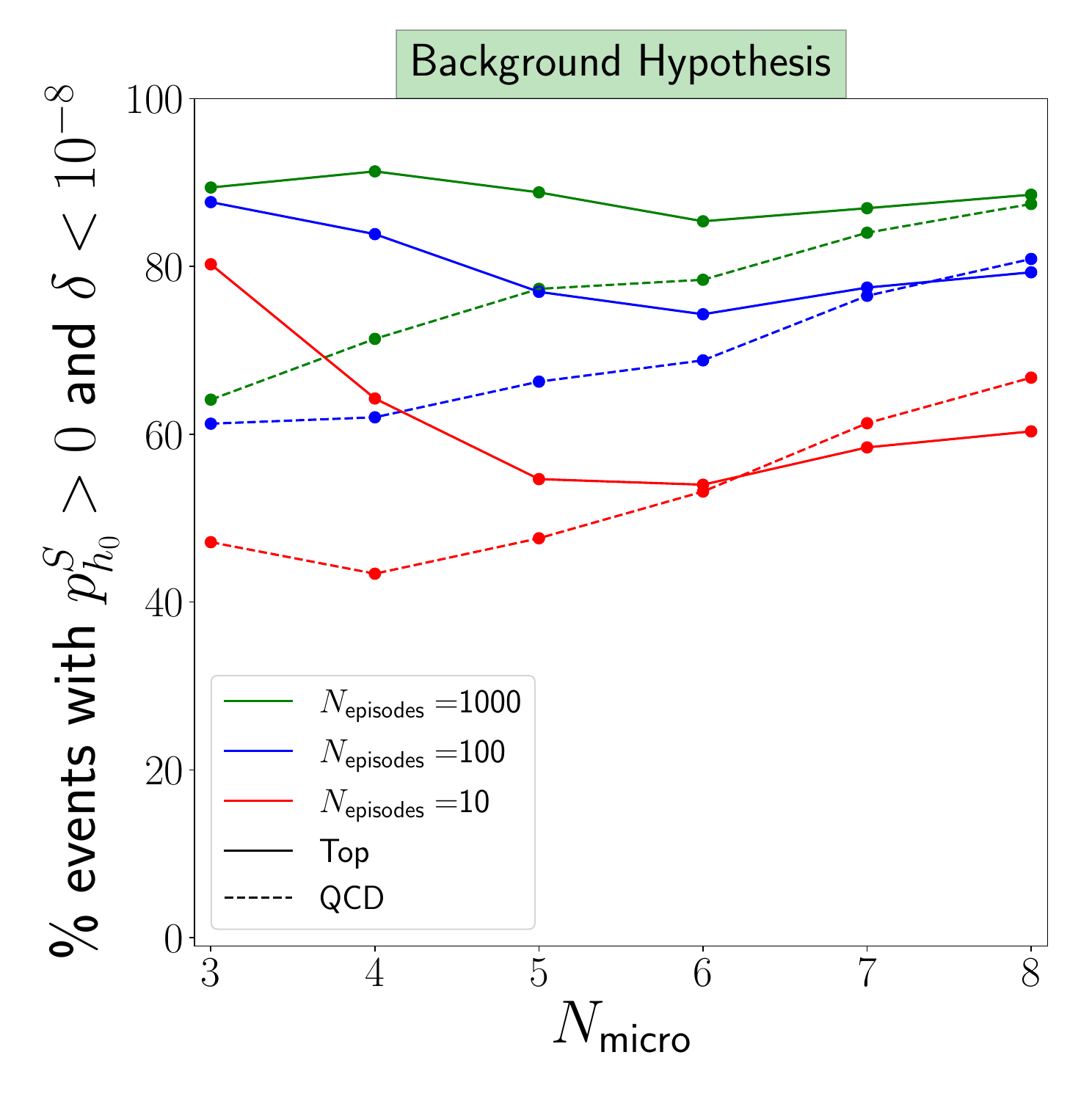}
	\includegraphics[scale=0.29]{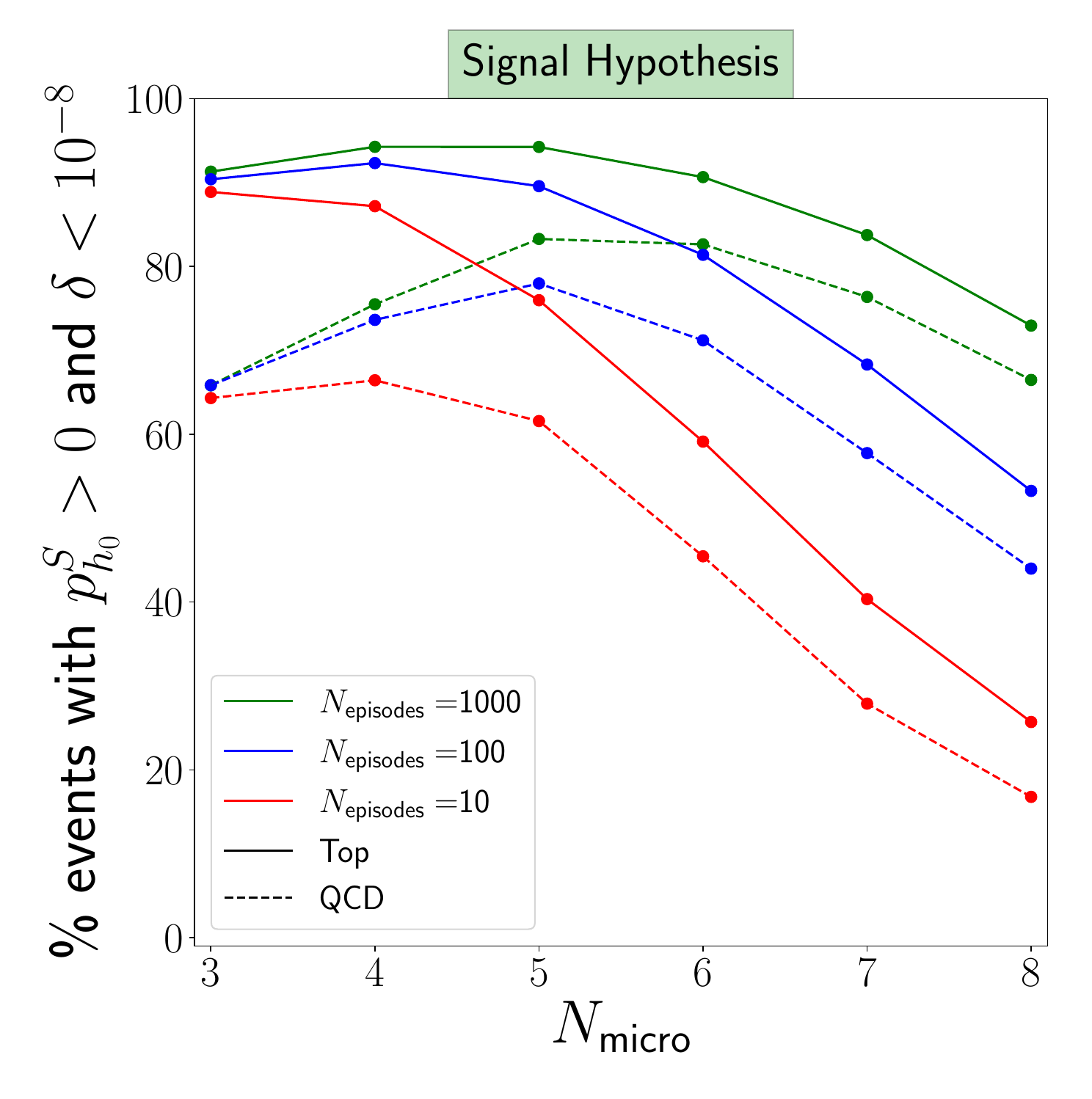}
	\caption{ We show the percentage of non-zero signal weighted jets with $\delta<10^{-8}$ varying with the microjet multiplicity $N_{\text{micro.}}$, for the signal (right) and background (left) hypothesis. }
	\label{fig:acc}
\end{figure}

\subsubsection{Comparisons to the optimal policy}
As AlphaPS aims to approximate the optimal ShowerMDP policy rather than finding the best possible discriminant, we first compare the accuracy of the approximated best shower history with the optimal one. Denoting the approximated history as $h_*$, the relative error of the approximation\footnote{Although $p_{h_0}\geq p_{h_*} $, we still define it with the absolute value as precision errors may numerically violate this condition.} is 
\begin{equation*}
	\delta=\frac{|p_{h_0}-p_{h_*}|}{p_{h_0}} 
\end{equation*}

The percentage of events\footnote{These percentages practically stays constant from $\delta<10^{-2}$ to $\delta<10^{-15}$ decreasing thereafter due to precision errors.} with $\delta<10^{-8}$ as a function of the microjet multiplicity of the fatjet $N_{\text{micro.}}$, for different number of episodes $N_{\text{episodes}}$, is shown in figure~\ref{fig:acc}. The relatively lower starting position for the QCD samples may be attributed to the relatively higher number of non-zero shower histories for the top dataset biasing the learning process. Due to this inherent bias, we will consider the features of the top datasets while discussing the overall features, as unbiased training should ideally result in the QCD sample's behaviour resembling that of the top samples. 
 Additionally, the neural agent does not explore all possible histories over different episodes as they are independent, dictated only by the neural agent's policy for each encountered state. A fallback to the brute-force approach should be the preferred option for the low multiplicity cases where the total shower histories are less than the $N_{\text{episodes}}$.    

For the signal hypothesis on the right, one sees a decreasing trend, which can be understood via the increasing search space complexity with increasing microjet multiplicity. The situation persists to a weaker extent for the background hypothesis on the left for $N_{\text{episode}}\in\{10,100\}$,  with the curve almost flat for $N_{\text{episodes}}=1000$. This trend can be understood from the relative differences in the distribution of non-zero weighted history for the background hypothesis compared to that of the signal. Imposing the Breit-Wigner shape on the top and W decays makes the shower weights fall relatively faster than the imposition of QCD splittings alone for the background hypothesis. Therefore, for the same amount of episodes, an agent can find shower histories which closely approximate the optimal shower history's weights and with $N_{\text{episodes}}=1000$, the neural agent has enough trials to have a very weak dependence on the scaling of the microjet multiplicity till $N_{\text{micro.}}=8$. It is important to reiterate that such a situation arises because the policy network has inductively learned the characteristics of the optimal policy, which reduces the sampling process to the vicinity of the highest weighted history.

\subsubsection{Discrimination with neural agent}

\begin{figure}[t!]
	\centering
	\includegraphics[scale=0.29]{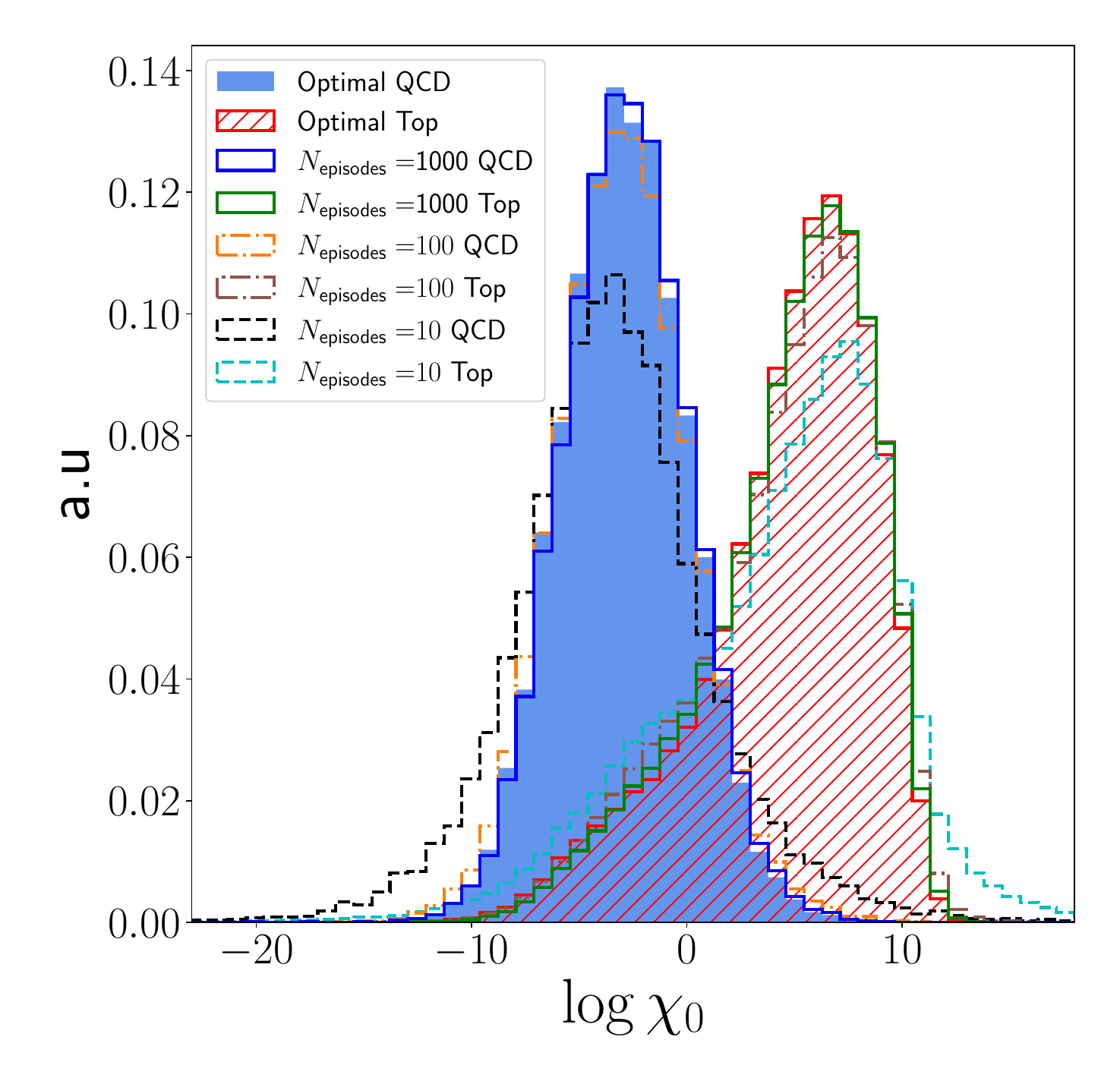}
	\includegraphics[scale=0.29]{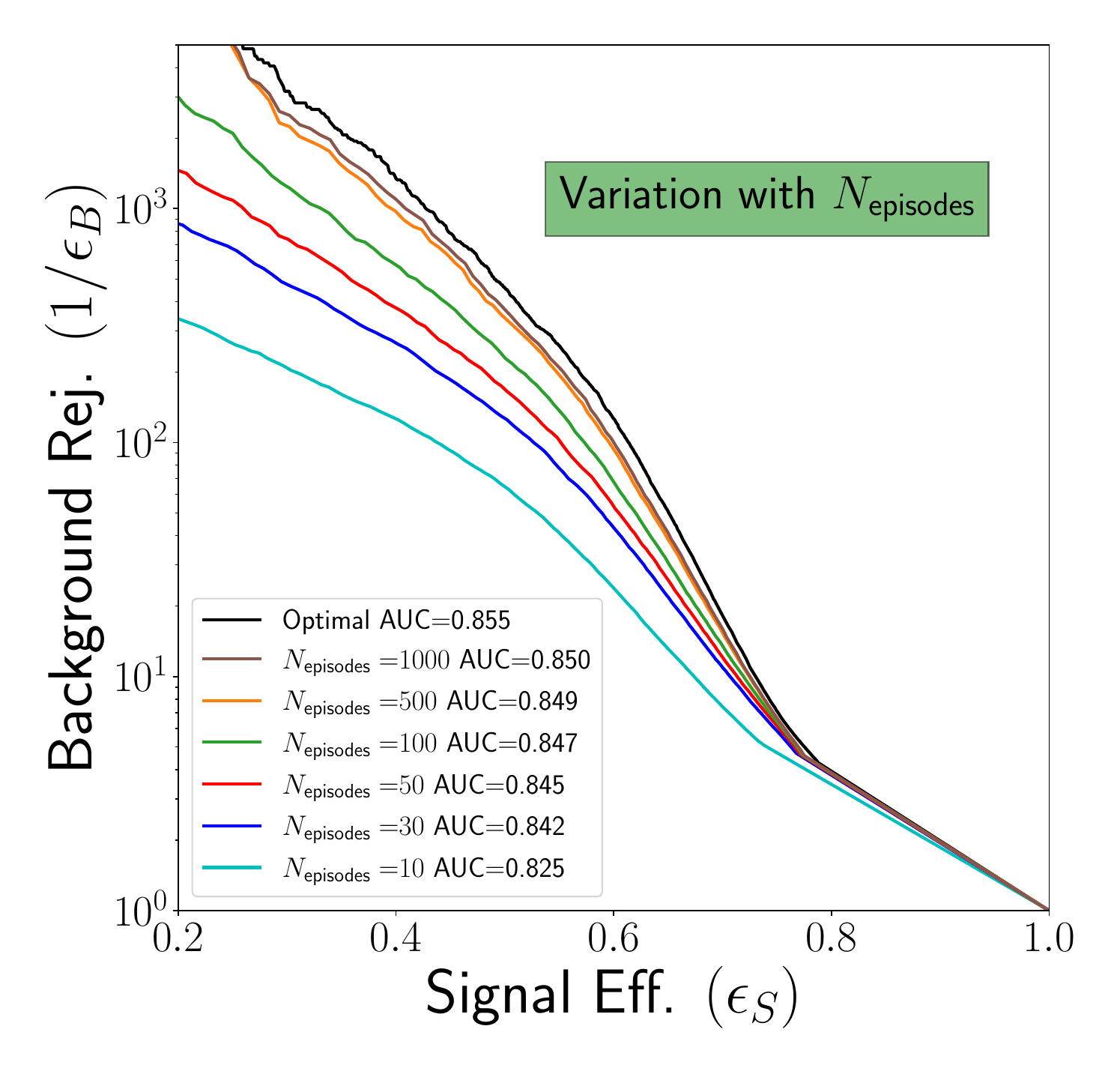}
	\caption{The figure shows the distribution of $\log\chi_0$ of the optimal history (after a brute force search), the neural network assisted history construction, and randomly constructed history of QCD and top jet samples. }
	\label{fig:chi_dist}
\end{figure}

Figure~\ref{fig:chi_dist} shows the distribution of $\log\chi_0$ evaluated with the best shower histories and the neural-assisted agent for different number of episodes $N_{\text{episodes}}$. The overlap between the optimally extracted $\log\chi_0$ increases with increasing $N_{\text{episodes}}$, pointing towards the agent learning to construct the individual shower histories correctly. The corresponding ROC curves and the AUC are shown on the right. As expected, the power of the constructed history increases with increasing $N_{\text{episodes}}$. The scaling of the AUC with increasing $N_{\text{episodes}}$ is relatively fast with $N_{\text{episodes}}=30$ giving an AUC of 0.842, which is around 1.5\% lower than the optimal AUC of 0.855. However, there is a limit to the possible discrimination due to the neural approximation as the AUC has already reached a plateau for $N_{\text{episodes}}=500$ with a slight increase to for $N_{\text{episodes}}=1000$.

\section{Conclusions}     
\label{sec:conc}

Our study presents a novel approach to the classification and inference of LHC data, addressing the computational challenges posed by the combinatorial explosion in the Shower Deconstruction method. By casting the problem of identifying the highest weighted shower history as a Markov Decision Process (MDP), we have demonstrated the practical application of a modular point-transformer architecture for learning the optimal policy. This neural agent, AlphaPS, has shown remarkable capability in constructing the highest weighted shower history, generalising well to unseen test data and significantly reducing complexity during inference.

The results from AlphaPS indicate a substantial improvement in computational efficiency, scaling linearly with constituent multiplicity. This significantly advances traditional matrix element methods, enabling practical applicability to high-multiplicity final states. Furthermore, our method maintains discrimination power comparable to the optimal ShowerMDP policy, showcasing the potential of deep learning techniques in particle physics data analysis in a physics-transparent approach.

This study opens several avenues for future research. Extending this framework to include a broader range of particle physics phenomena could provide deeper insights into LHC data interpretation. Secondly, further refinement of the neural network architecture and training process with exploration-based reinforcement learning could yield more efficient, accurate models and generalisations to higher multiplicities. Lastly, exploring the integration of this approach with other machine-learning techniques could lead to a more holistic and powerful tool for particle physics data analysis.

We demonstrate the promising synergy between deep learning and particle physics, offering a novel perspective in interpreting complex LHC data and setting a precedent for future explorations in this interdisciplinary field.

\section*{Acknowledgements}
M.S. and V. S. N. are supported by the STFC under grant ST/P001246/1. Computational work were performed on the Param Vikram-1000 High Performance Computing Cluster and TDP resources at the Physical Research Laboratory (PRL).

\appendix 

\section{Details of Network architecture and training} 
\label{app:arch} 
In the appendix, we detail the network architecture and the training procedure for the neural function approximation of the optimal policy. Independent of the input and output dimensions, all multilayer-perceptrons (MLPs) in our architecture have two hidden layers of 128 nodes and ReLU activation functions. All networks are implemented and trained using \textsc{PyTorch-Geometric}~\cite{Fey/Lenssen/2019}
\subsection{Value Network}
The value network determines whether a given fatjet, with its microjet set, should undergo the construction of shower histories based on the finiteness of its signal weight.  It's aim  is to learn a function from the initial microjet set  to the interval $(-1,1)$, where negative values signify that the jet has only zero-weighted signal histories, while positive values signify that there is at least one with a finite weight. Since the task of identifying microjet sets with no combinatorial divisions in the top or $W$ mass window (thereby having only zero weighted histories) is relatively easier than understanding the evolution of the microjets based on perturbative QCD splittings, we employ a Deep Sets~\cite{NIPS2017_f22e4747} architecture for the value network. Mathematically, it can be written as
\begin{equation}
	\hat{v}(\{\mathbf{h}_1,\mathbf{h_2},...,\mathbf{h}_{N_\text{micro}}\})=\Phi\left(\frac{1}{N_{\text{micro}}}\sum_{i=1}^{N_{\text{micro}}}\; \phi(\mathbf{h}_i)  \right) 	
\end{equation}

It consists of two MLPs: $\phi$ and $\Phi$. The first one $\phi$, map each microjet feature $\mathbf{h}_i$ (given in Eq.~\ref{eq:micro_state}) to a 128-dimensional latent node representation $\phi(\mathbf{h}_i)$. We perform a mean readout of these node features, which feeds to $\Phi$. The output of this second MLP is made to fall within $(-1,1)$ via a hyperbolic tangent activation function.

\subsection{Policy Network}
The flavour policy modules have two distinct parts: a message-passing block and downstream processing of the message-passing block's output specific to the nature of the action space. In the following passage, we discuss these two stages separately. 
\paragraph{Message Passing Blocks:} Two message-passing-neural-network(MPNN)-blocks are common for the policy for all flavours except the root edge, one each for the microjet state (micro-MPNN) and the tree state (tree-MPNN). A flavor-embedding MultiLayer Perceptron (FE-MLP), which has two hidden layers of thirty-two nodes and ReLU activation, takes the PIDs of the tree state's nodes and embeds them into a four-dimensional vector. The concatenated vector formed by the tree state's node features (Eq~\ref{eq:micro_state}) and the four-dimensional flavour embedding feeds as the node features to the tree-MPNN. The policy for the root edge has a single MPNN for the microjet state.

The MPNN blocks are based on the \textsc{PointTransformerConv}~\cite{Zhao_2021_ICCV}, which automatically learns vector attention for each node with every other node in the graph. This attention mechanism is advantageous for the microjet set, where, without any input in the graph construction, the network automatically learns the interrelations needed for the division of the microjets. The micro-MPNN and the tree-MPNN have two message-passing operations for all relevant flavours and hypotheses except the top quark's micro-MPNN and the micro-MPNN of the gluon flavour module for the background hypothesis, where we found an improved accuracy when utilizing three message-passing operations. Due to the computational demands of increasing message-passing operations, we did not consider more operations for other flavours.  

The point transformer operation updates the node features $\mathbf{h}_i^{(l)}$ at a message-passing stage $l$ to 
\begin{equation}
	\mathbf{h}_{i}^{(l+1)}=\sum_{j\in\mathcal{N}(i)}\, \mathbf{w}_{ij}\,.\,\left( \alpha(\mathbf{h}^{(l)}_j)+\delta(\mathbf{p}_i-\mathbf{p}_j)\right) \quad,
\end{equation}where  $\mathcal{N}(i)$ is the neighbourhood of the node $i$, $\mathbf{w}_{ij}$ are the attentions weights, $\alpha$ is a linear layer, and $\mathbf{p}_i$ and $\mathbf{p}_j$ are the coordinate inputs for the position MLP $\delta$. The weights $\mathbf{w}_{ij}$ are calculated as 
\begin{equation}
	\mathbf{w}_{ij}=\rho\left(\gamma(\,\phi(\mathbf{h}^{(l)}_i)-\psi(\mathbf{h}^{(l)}_j)+\delta (\mathbf{p}_i-\mathbf{p}_j) \, )\right)\quad, 
\end{equation}with $\phi$ and $\psi$ denoting linear layers, and $\gamma$ denoting our choice of mapping function. $\rho$ is a \textsc{SoftMax} normalization over the $j$ axis. All point-transformer convolutions in the policy and the value network outputs a 128-dimensional node feature. The $\delta$ MLPs take the input node-features' two-dimensional rapidity-azimuth information and map it to the 128-dimensional output feature space. For $\gamma$, we use MLPs that map the 128-dimensional vector in the argument to a 128-dimensional output node feature space through which the attention weights are calculated for each feature dimension separately. To enable the downstream network to segregate information at different scales, we take the output of each message-passing operation and form a concatenated $L\times 128$ dimensional node feature vector as the output of an MPNN block, with $L$ being the number of message-passing operations.

 If present, the output node features of the tree-MPNN undergo a mean graph readout. This graph representation gets concatenated with each node feature output of the micro-MPNN, which acts as the input to all the downstream networks.
 
\paragraph{Downstream Architecture:} For all flavours, the policy network has a microjet-assignment-MLP (MA-MLP), which takes the combined node representation and outputs a single value in $(0,1)$ as the node's set partitioning probability. We implement a sigmoidal activation for the output of this MLP to have a probabilistic interpretation. We have two additional MLPs for the flavour assignment of the gluon and top quark. The first MLP takes the combined representation and outputs a node-level feature embedding. We perform a mean graph readout of the node representations and feed it into the second  MLP, which outputs a fixed dimensional array of $2$ or $3$-dimensional matrix depending on the flavour. These matrices undergo a softmax normalization for the probabilistic interpretation. 
\subsection{Training}
\begin{table*}[t!]
	\centering 
			\begin{tabular}{ccc}\hline
					 \textbf{Flavor} && \textbf{Best Accuracy} \\
					\hline   			
					\multicolumn{1}{l}{\textit{Value}}&&\\ 
					\hline 
					NA & & 0.9708\\
					\hline
					\multicolumn{1}{l}{\textit{Signal Hypothesis}}&&\\ 
					\hline 
					Root && 0.8274\\
					Top && 0.6638\\ 
					$W$ && 0.7575\\
					Gluon && 0.9077\\ 
					Bottom && 0.8863\\
					Quarks && 0.9357\\
					\hline
					\multicolumn{1}{l}{\textit{Background Hypothesis}}&&\\
					\hline 
					Root & &0.5533\\
					Gluon &&0.7513\\
					Quarks &&0.9057\\
					\hline 
				\end{tabular}
	\caption{The table shows the accuracy of the best epoch on the validation dataset for the value network and the action prediction of the flavour policy modules.}
	\label{tab:percentage_acc}
\end{table*}

All networks are trained on the extracted data independently for each flavor-hypothesis pair for a hundred epochs using the \texttt{Adam} optimizer~\cite{DBLP:journals/corr/KingmaB14} initialized with a learning rate of 0.001. The learning rate decreases by half if the validation loss has not decreased for three epochs. We use the binary and categorical cross-entropy as the loss function for the microjet and flavour assignment actions, respectively, while for the value network, we use mean squared error. We use the epoch with the minimum validation loss for inference.  

We can estimate the learned understanding for the different modules by extracting the accuracy of the predicted value, the predicted set partitioning, and flavour assignment actions on the validation dataset. For the value, this is done by assigning -1 to negative outputs and +1 to positive outputs. We convert the fuzzy logical output of a sigmoid set partitioning probability $p$ via the boolean expression $p>0.5$. At the same time, for the split assignment action, we transform the softmax output vectors to one-hot encoded vectors by assigning one to the index with the largest entry.

The accuracy of the best epoch on the validation dataset for the various modules is shown in Table~\ref{tab:percentage_acc}. Due to the relative ease of identifying jets falling outside the top and $W$ mass window, the value network has a high accuracy of 0.9604. The relative accuracy of the various policy modules helps estimate the difficulty level in learning the optimal actions for the respective flavour-hypothesis pair. For each hypothesis, the flavour of the first FSR set, the top quark for the signal, and the gluon for the background have the least accuracy. To understand this, we note that the action choice for these FSR flavours depends on the tree's full depth, which would be more complex than other flavours. Consequently, the signal hypothesis's gluon accuracy is relatively much better due to its occurrence at later stages of the shower tree than the top quark. For all other flavours, the best accuracy is somewhat similar, pointing towards the easier choice of optimal actions due in part to the lower diversity of the encountered edges further down the shower history.

Note that the accuracy of the root edge is much better than the FSR edge's flavour for the signal hypothesis, which can be understood if we recall that the larger multiplicity of the top decay demands more microjets in the FSR sets, with the highest factor mainly arising from the initial set's constituents going to the FSR set, which is easily learned by the policy module. On the other hand, the division of the background root edge into ISR and FSR subsets is highly non-trivial, and the root edge's accuracy is not as good as that of the signal. This accuracy, being close to 50\%, is a significant improvement from a randomly acting agent whose accuracy scales as the inverse of the number of possible combinatorial divisions of the microjet set into two distinct subsets.

\bibliographystyle{JHEP}
\bibliography{ref.bib}
	
\end{document}